\documentclass[aps,prl,twocolumn,amsmath,amssymb,superscriptaddress,hidelinks]{revtex4-2} 
\usepackage{bm}
\usepackage{verbatim} 
\usepackage{graphicx}
\usepackage[usenames,dvipsnames]{color}
\usepackage[colorlinks]{hyperref}
\hypersetup{
  colorlinks,
  citecolor=blue,
  linkcolor=blue,
  urlcolor=blue}
\usepackage{natbib}
\usepackage{array}[=2016-10-06]
\usepackage{tabularx}
\usepackage{booktabs}

\newcommand{\be}{\begin{equation}}
\newcommand{\ee}{\end{equation}}
\newcommand{\bea}{\begin{eqnarray}}
\newcommand{\eea}{\end{eqnarray}}

\newcommand{\p}{\partial}

\newcommand{\la}{\langle}
\newcommand{\ra}{\rangle}

\newcommand{\Tr}{{\rm \, Tr\,}}

\renewcommand{\Im}{{\rm Im}\,}
\renewcommand{\Re}{{\rm Re}\,}
\renewcommand{\vec}[1]{{\bf #1}}
\newcommand{\dd}{\mathrm{d}}
\newcommand{\BZ}{\mathrm{BZ}}

\newcommand{\OO}[1]{\mathcal O(#1)}

\newcommand{\avgib}[1]{\left\langle #1\right\rangle_{\!\mathrm{ib}}}
\newcommand{\avgn}[1]{\left\langle #1\right\rangle_{\!n}}

\newcommand{\todo}[1]{\textcolor{magenta}{#1}}

\newcolumntype{Y}{>{\centering\arraybackslash}X}

\begin{document}

\title{Nonlinear spin generation and spin photocurrent as a symmetry and magnetic domain probe in unconventional magnets}

\author{Ying Xiong}
\affiliation{Science, Mathematics and Technology, Singapore University of Technology and Design, Singapore 487372}
\author{Er-Ping Li}
\affiliation{College of Information Science and Electronic Engineering, Zhejiang University, Hangzhou 310027, China}
\author{L. K. Ang}
\email{ricky\_ang@sutd.edu.sg}
\affiliation{Science, Mathematics and Technology, Singapore University of Technology and Design, Singapore 487372}

\begin{abstract}
Unconventional magnets host momentum-dependent spin splitting without spin–orbit coupling. We demonstrate that the combination of nonlinear spin polarisation and spin photocurrent is a powerful diagnostic tool for probing the spin space group symmetry, magnetic domain, and quantum geometry in unconventional magnets, providing information beyond their equilibrium spin texture. We develop a mechanism-resolved joint symmetry analysis of second-order spin photoresponse tensors and analyse their frequency dependence, magnetic parity and relaxation scaling. We show that unconventional magnets are natural platforms to achieve large magnetic-even spin photocurrent induced by unpolarised light, making them promising candidates for polariser-free spin sources. We present numerical illustrations for prototypical hybrid parity and odd parity magnet models. Our work provides optical fingerprints connecting spin-space symmetry to distinct microscopic pathways of spin accumulation and transport.
\end{abstract}

\maketitle

\begin{figure}[t!]
 \includegraphics[width=0.95\columnwidth]{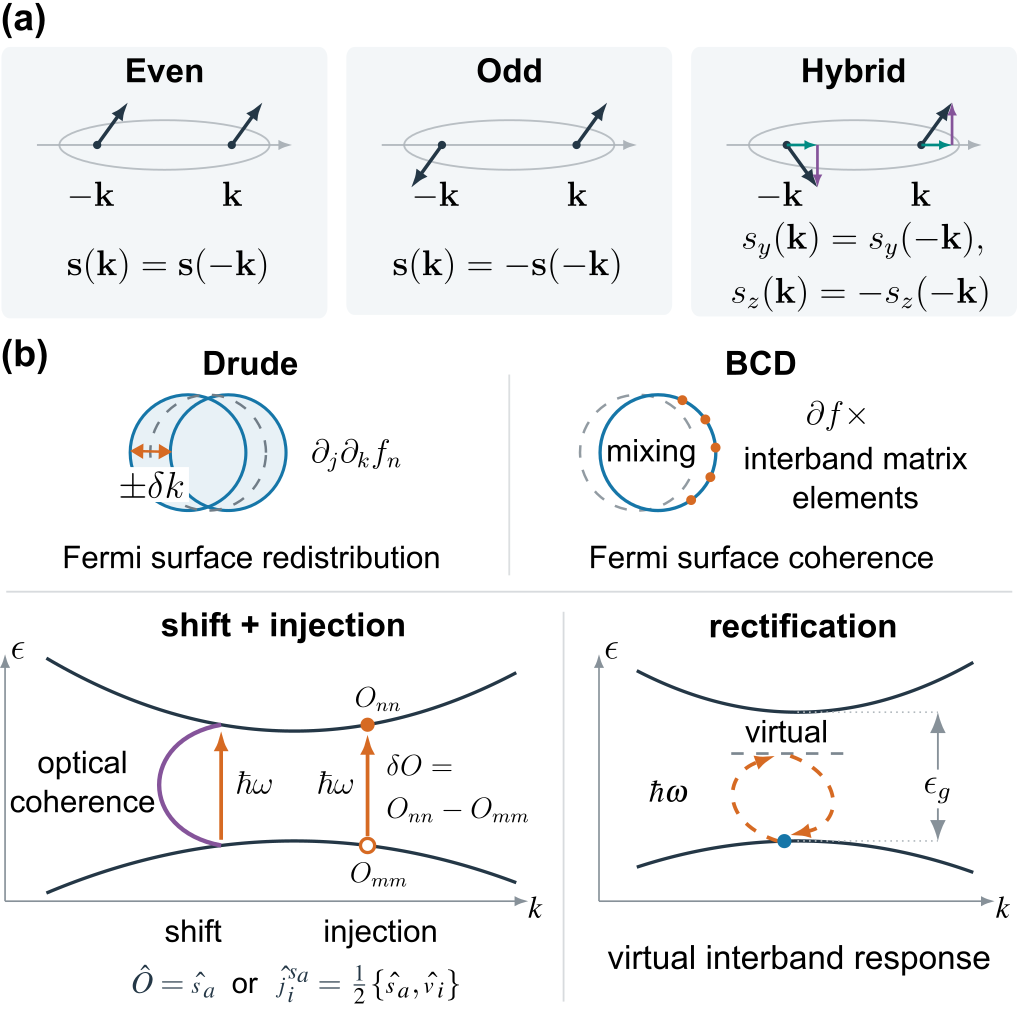}
 \caption{(a) Even, odd, and hybrid parity spin textures in momentum space. (b) Microscopic mechanisms for second order responses of $\hat O=\hat s_a$ or $\hat j_i^{s_a}$. The semiclassical Drude contribution arises from Fermi surface redistribution; BCD combines the Fermi surface gradient with interband mixing. Injection and shift contributions are interband resonant responses induced by a change in the expectation value $\delta O=O_{nn}-O_{mm}$ of the corresponding observables and interband coherence, respectively. Rectification contribution arises from virtual interband processes and can be nonzero even when $\hbar \omega$ is smaller than the band gap.}
 \label{Fig:schematic}
\end{figure}

Unconventional magnets have attracted great attention because compensated magnetic order can coexist with momentum-dependent spin splitting without spin--orbit coupling~\cite{Smejkal2022,Hayami2019,
Smejkal2022Landscape,Jungwirth2026Signatures,Krempasky2024,Yu2025, Song2025, Luo2026}. They can be classified by spin parity in momentum space: nonzero components of the Bloch-spin texture are even [$\vec s(\vec k) = \vec s (-\vec k)$], odd [$\vec s(\vec k) = -\vec s (-\vec k)$]~\cite{Yu2025, Song2025}, or of hybrid parities [$s_a(\vec k) = s_a (-\vec k)$, $s_b(\vec k) = - s_b (-\vec k)$]~\cite{Luo2026, YeeSin2026}, as illustrated in Fig.~\ref{Fig:schematic}a. Prominent examples include altermagnets which exhibit non-relativistic alternating spin splitting in momentum space~\cite{Smejkal2022,Hayami2019,
Smejkal2022Landscape,Jungwirth2026Signatures,Krempasky2024, Cheong2025,Song2025Functional, Wang2026} and the newly emerging hybrid parity magnets~\cite{Luo2026}. These diverse emerging unconventional magnets offer new possibilities for spintronic applications such as spin generation, transport, and memory devices~\cite{Song2025Functional, Luo2026, Naka2019, GonzalezHernandez2021, Chakraborty2025, Mencos2026,Jungwirth2026Spintronics, Smejkal2022GMR}, while creating a diagnostic challenge: distinct spin symmetries and magnetic domains can share the same equilibrium spin texture and zero net magnetization~\cite{Chen2024}.

Nonlinear optical responses are sensitive to quantum geometry of the Bloch wavefunctions~\cite{deJuan2017, Sodemann2015, Watanabe2021, Ahn2020, Xiong2022, Ma2023, Gao2026review}. In particular, nonlinear spin polarisation and spin photocurrents are powerful tools for probing the spin space group symmetry, spin-dependent interband coherence and quantum geometric properties of the spinor Bloch states in magnetic materials~\cite{Tarasenko2005, Lihm2022,Sarkar2024, Xu2021,Xiao2022, Takagi2026, Feng2025, Feng2026}. 
Recent work has established spin-symmetry constraints on spin photocurrents in unconventional magnets, alongside the connection between magnetic order, quantum geometry, and nonlinear transport~\cite{Xiao2026, Dong2025, Sivianes2025}. These advances motivate a joint diagnostic tool based on nonlinear spin accumulation and spin photocurrent. Their different symmetry transformations offer complementary information, but exploiting it requires identifying which optical channels distinguish magnetic symmetries, which microscopic processes they select, and their parity in magnetic order parameter.

\begin{table*}
\centering
\begin{tabularx}{\textwidth}{Y Y Y Y Y Y Y Y Y }
\hline
\hline
 & $U_{nz}$ & $\mathcal T U_{2z} P$ & $\mathcal T U_{2z}$ &  $\mathcal T U_{2x}$ & $U_{2y} P$ & $U_{2z} P$ & $\mathcal T$ & $P$ \\  
$\chi^{s_x}_{jk}$ & $\times$ & O & O& E& $\times$ & $\times$ & E & E/O \\
$\chi^{s_y}_{jk}$ & $\times$ & O & O & O & E/O & $\times$ & E & E/O \\
$\chi^{s_z}_{jk}$ & E/O & E &E & O & $\times$ & E/O & E & E/O \\
$\sigma^{s_x}_{i,jk}$ & $\times$ & E & O & E& E/O & E/O & E & $\times$ \\
$\sigma^{s_y}_{i,jk}$ & $\times$ & E & O & O & $\times$ & E/O & E & $\times$ \\
$\sigma^{s_z}_{i,jk}$ & E/O & O & E & O & E/O & $\times$ & E & $\times$ \\ 
\hline 
\hline
\end{tabularx}
\caption{Symmetry constraints on second-order spin generation and spin photocurrent tensors by combinations of spin rotation, time reversal and spatial inversion symmetries. E (O) indicates if the allowed response is even or odd under time reversal, and $\times$ indicates that the response is forbidden. }
\label{tab:macroscopic_sym}
\end{table*}

We provide a systematic symmetry and mechanism analysis for second order spin polarisation and spin photocurrent in unconventional magnets, with emphasis on the newly emerging odd- and hybrid-parity magnets~\cite{Yu2025,Song2025, Luo2026, YeeSin2026}. A joint, mechanism-resolved treatment of spin polarisation and spin current identifies complementary optical channels in which competing nonlinear processes are eliminated by symmetry or suppressed in a controlled relaxation limit. We present numerical illustrations of the optical selection rules in prototypical hybrid-parity and odd-parity (see End Matter) magnet models. Furthermore, we show that the symmetry-allowed unpolarised response suggests a route to polariser-free spin sources with distinct microscopic origins and magnetic-domain parities. Owing to the symmetry constraints, we isolate the domain-odd and domain-even responses: the former provides an all-optical probe for the magnetic domain of an unconventional magnet, while the latter enables a potentially large response across magnetic domains.

\textcolor{blue}{\it Symmetry-constrained nonlinear spin magnetization and spin photocurrent.}
Light irradiation can induce nonlinear spin magnetization and spin photocurrent which are constrained by symmetries. Consider an AC electric field $\vec E (t) = \frac{1}{2} \vec E e^{-i \omega t} + c.c $, the time independent second order spin generation can be written as $s_a^{(2)} = \sum_{j, k} \chi^{s_a}_{jk} E_j E_k^*$, and the DC spin photocurrent flowing along the $i$ direction can be written as $j^{(2), s_a}_i = \sum_{j, k} \sigma^{s_a}_{i,jk} E_j E_k^*$. 
Let $\mathcal M$ denote the complete magnetic order and define magnetic-even/odd sectors of a nonlinear tensor $K \in \{ \chi, \sigma\}$ by $K^{(E/O)}[\mathcal M]=\{K[\mathcal M]\pm K[-\mathcal M]\}/2$ at the same positive damping. For a preserved operation $g=\{\mathcal T^{t_g}U\Vert R|\bm t\}$, with proper spin rotation $U$, spatial operation $R$, time reversal $\mathcal T$, and $t_g=0,1$, the constraints are
\begin{align}
 \chi^{s_a,(p)}_{jk}&=p^{t_g}U_{ab}R_{jq}R_{kr}\chi^{s_b,(p)}_{qr},\label{eq:chi_sym}\\
 \sigma^{s_a,(p)}_{i,jk}&=p^{t_g}U_{ab}R_{i\ell}R_{jq}R_{kr}\sigma^{s_b,(p)}_{\ell,qr},\label{eq:sigma_sym}
\end{align}
where $p=+1$ ($E$) or $-1$ ($O$), and repeated indices are summed. For an antiunitary symmetry its unitary part maps $\mathcal M$ to $-\mathcal M$, producing the factor $p$. Here, $U$ and $R$ act independently on the spin and real space degrees of freedom. 

We summarize the constraints on the spin polarisation tensor $\chi^{s_a}_{jk}$ and spin photocurrent tensor $\sigma^{s_a}_{i,jk}$ by a range of symmetries involving $\mathcal T$, spatial inversion $P$ and spin rotations, see Table~\ref{tab:macroscopic_sym}. Further lattice symmetries acting on real space such as mirror reflection can further limit the responses associated with light polarisation and spin photocurrent flow directions. We have also suppressed the translations here since they do not affect response-tensor in a homogeneous system.
The nonlinear response tensors depend on the full spin space symmetries and not just the spin parity in $k$-space. Importantly, the spin polarisation and spin photocurrent response tensors exhibit contrasting symmetry selection rules with respect to time reversal and spatial inversion: $\hat s_a$ is invariant under $P$ but changes sign under $\mathcal T$, while $\hat j^{s_a}_i$ changes sign under $P$ but is invariant under $\mathcal T$. 

The combination of $\chi^{s_a}_{jk}$ and $\sigma^{s_a}_{i,jk}$ enables to probe the spin space group of the material even when the equilibrium macroscopic magnetization are indistinguishable. For example, both the collinear Type I odd parity magnet (preserving $U_{nz}$ and $\mathcal T U_{2z}$, with $s_z(\vec k) = -s_z(-\vec k), \; s_{x,y}(\vec k) =0$) and the coplanar Type II odd parity magnet (preserving $\mathcal T U_{2z} P$ and $\mathcal T$ with $s_{x,y}(\vec k) = -s_{x,y} (-\vec k), \; s_z(\vec k) =0$) support zero equilibrium magnetization and only nonzero $\chi^{s_z, (E)}_{jk}$ despite their contrasting spin texture. However, their spin group symmetries can be further distinguished by measuring the spin photocurrent tensors, with nonzero $\sigma^{s_z, (E)}_{i,jk}$ for the former, and $\sigma^{s_{x,y}, (E)}_{i,jk}$ for the latter.  This enables complementary optical fingerprints of spin group symmetry.

Furthermore, magnetic odd nonlinear responses are also powerful tools to detect magnetic domains without actually flipping the magnetic order~\cite{Merte2023, Pei2026, Rappe2026, Bai2026}. This is particularly useful for fully compensated unconventional magnets, where the equilibrium magnetization is zero. For example, collinear odd-parity magnet preserving $U_{nz}$ and $U_{2y} P$ with $s_z (\vec k) = -s_z (-\vec k)$ and $s_{x,y} (\vec k) = 0$ supports both magnetic even and odd $\sigma^{s_z}_{i,jk}$. Magnetic domains can be distinguished by comparing the sign and magnitude of the spin photocurrent measurement.

\begin{table*}[t]
\centering
\renewcommand{\arraystretch}{2.3}
\begin{tabularx}{\textwidth}{l >{\centering\arraybackslash}X l}
\toprule
 & Contribution to $K^O_{jk}$ & Physical content\\
\midrule
Drude & $\displaystyle\frac{e^2}{4(\Omega^2+\eta^2)}\avgn{O_{nn}\partial_j\partial_kf_n}$ & Fermi-surface redistribution\\
BCD & $\displaystyle-\frac{ie^2}{4}\avgib{\frac{O_{nm}}{\Lambda_{mn}}\left(\frac{R^j_{mn}\partial_kF_{mn}}{d_+}+\frac{R^k_{mn}\partial_jF_{mn}}{d_-}\right)}$ & Fermi-surface coherence\\
Injection & $\displaystyle\frac{\pi e^2\tau^{\rm inj}}{2\hbar}\avgib{W^O_{jk,mn}\delta(X_{mn})}$ & Resonant population transfer\\
Shift & $\displaystyle-\frac{i\pi e^2}{4\hbar}\avgib{S^O_{jk,mn}\delta(X_{mn})}$ & Resonant interband coherence\\
Rectification & $\displaystyle-\frac{e^2}{4\hbar}\avgib{G^O_{jk,mn}(\eta)P_\eta(X_{mn})}+\frac{e^2}{4}\avgib{W^O_{jk,mn}H_\eta(X_{mn})}$ & Virtual interband processes\\
\bottomrule
\end{tabularx}
\caption{Microscopic contributions to $\chi^{s_a}_{jk}$ and $\sigma^{s_a}_{i,jk}$. $\la A_n \ra_n = \int_{\rm BZ} d\vec k/(2\pi)^d \sum_n A_n$ indicates the integration of the Brillouin zone and summing over all bands, $\avgib{A_{mn}}=\int_\BZ d\vec k /(2\pi)^{d}\sum_{m\ne n}A_{mn}$ and $d$ is the dimension of the material. Here, $\hat O=\hat s_a$ for $\chi^{s_a}_{jk}$ or $\hat O=\hat j^{s_a}_i$ for $\sigma^{s_a}_{i,jk}$. See the End Matter for expressions of the matrix elements and derivations.}
\label{tab:microscopic}
\end{table*}

\textcolor{blue}{\it Optical selection of microscopic mechanisms and the clean limit.} Not only can symmetry constrain the total light-induced spin polarisation and spin photocurrent, it also enables us to selectively access its different microscopic origins, which serve as a powerful tool to probe band gap and quantum geometric properties (in combined reciprocal and spin space). For either measured operator $\hat O=\hat s_a$ or $\hat j_i^{s_a} = \{ \hat s_a, \hat v_i \}/2$, we group the microscopic origins into three categories\cite{Lihm2022}: the Fermi surface contributions [Drude and Berry curvature dipole (BCD)], direct optical transitions (injection and shift), and virtual interband transitions (rectification). We have 
\begin{equation}
 K=K^{\mathrm D}+K^{\mathrm{BCD}}+K^{\mathrm{inj}}
       +K^{\mathrm{shift}}+K^{\mathrm{rec}},\label{eq:five}
\end{equation}
where their expressions are summarised in Table~\ref{tab:microscopic}. 

The semiclassical Drude contribution arises from a second-order occupation redistribution at the Fermi surface. The BCD-like sector combines an occupation gradient with interband matrix elements and reduces to the familiar Berry-curvature-dipole structure analogous to that of charge photocurrent~\cite{Sodemann2015,Lihm2022,Xiao2026}. Injection contribution measures the difference between the final- and initial-state spin or spin current expectation values, multiplied by an optical transition rate and a population accumulation time $\tau^{\rm inj}$. Here, $\tau^{\rm inj}$ depends on the pulse width as well as the relaxation time of photoexcited spin imbalance or spin photocurrent~\cite{deJuan2017, Holder2020, Tarasenko2005}, see End Matter for a detailed discussion. Shift contribution arises from the off-diagonal components of the nonequilibrium density matrix during optical transitions and contains the interband coherence.
Rectification arises from virtual interband processes. These kernels combine optical Berry-connection matrix elements with spin-dependent observables, providing access to wavefunction quantum geometric properties beyond the band dispersion~\cite{Lihm2022,Sarkar2024, Xu2021, Ahn2020}.

\begin{table*}
\centering
\begin{tabularx}{\textwidth}{Y Y Y Y Y Y Y Y Y }
\hline
\hline
mechanism & $\Re \chi^{(e)}$ & $\Re \chi^{(o)}$ & $\Im \chi^{(e)}$ &$\Im \chi^{(o)}$ & $\Re \sigma^{(e)}$ & $\Re \sigma^{(o)}$ & $\Im \sigma^{(e)}$ &$\Im \sigma^{(o)}$ \\
Drude & $\times$ & $\mathcal O(1)$ & $\times$ & $\times$ & $\mathcal O(1)$ & $\times$ & $\times$ & $\times$ \\
BCD & $\mathcal O(\eta)$ & $\mathcal O(\eta^2)$ & $\mathcal O(1)$ & $\mathcal O(\eta)$ & $\mathcal O (\eta^2)$ & $\mathcal O(\eta)$ & $\mathcal O(\eta)$ & $\mathcal O(1)$ \\
rectification & $\mathcal O(\eta)$ & $\mathcal O(1)$ & $\mathcal O(1)$ & $\mathcal O(\eta)$ & $\mathcal O(1)$ & $\mathcal O(\eta)$ & $\mathcal O(\eta)$& $\mathcal O(1)$ \\
injection & $\times$ & $\mathcal O(1)$ & $\mathcal O(1)$ & $\times $ & $\mathcal O(1)$ & $\times $ & $\times$ & $\mathcal O(1)$ \\
shift & $\mathcal O(1)$ & $\times $ & $\times$ & $\mathcal O(1)$ & $\times$ & $\mathcal O(1)$ & $\mathcal O(1)$ & $\times $ \\
\hline 
\hline
\end{tabularx}
\caption{The scaling of relaxation rate $\eta$ for each contribution of the nonlinear spin polarisation and spin photocurrent tensor for a fixed, nonzero optical frequency. }
\label{tab:eta_scale}
\end{table*}

As we now discuss, different microscopic contributions can be distinguished by symmetry constraints and dependence on light frequency as well as dissipation/relaxation rate $\eta$ of the material. We separate each microscopic contribution into components even and odd under reversing of magnetic order $\mathcal M$, and the real and imaginary parts of each tensor denote responses that depend on the linear polarisation and helicity of the incident light respectively. Table~\ref{tab:eta_scale} summarises the scaling of each contribution as a function of $\eta$. As we can see, while the Drude contribution is always real and odd (even) for spin polarisation (photocurrent) response, other mechanisms generally support both magnetic odd and even responses, and their linear vs helicity (real vs imaginary) dependence and $\eta$ dependence is determined by the role of energy absorption and dissipation.

Combining with the symmetry constraints in Tables~\ref{tab:macroscopic_sym} and~\ref{tab:eta_scale}, this allows us to filter out specific microscopic mechanisms and probe their associated quantum geometry by optical control. As we see, for the hybrid parity magnet model illustrated below, the helicity-dependent spin-polarization channel selects the combined resonant shift contribution in the regular clean absorbing limit. Drude and injection do not contribute to its imaginary part, while the BCD and rectification corrections disappear as $\eta \to 0$. Conversely, the spin-current analysis provides complementary clean-limit shift channels: the real $s_x$-polarized channels and the imaginary $s_z$-polarized channel select resonant shift contributions in the clean limit. In the following, we show numerical illustration of these optical selections rules in a hybrid-parity magnet model, a complementary example for a newly emerging odd-parity magnet CeNiAsO is presented in the End Matter.

\begin{figure}[t]
  \includegraphics[width=\linewidth]{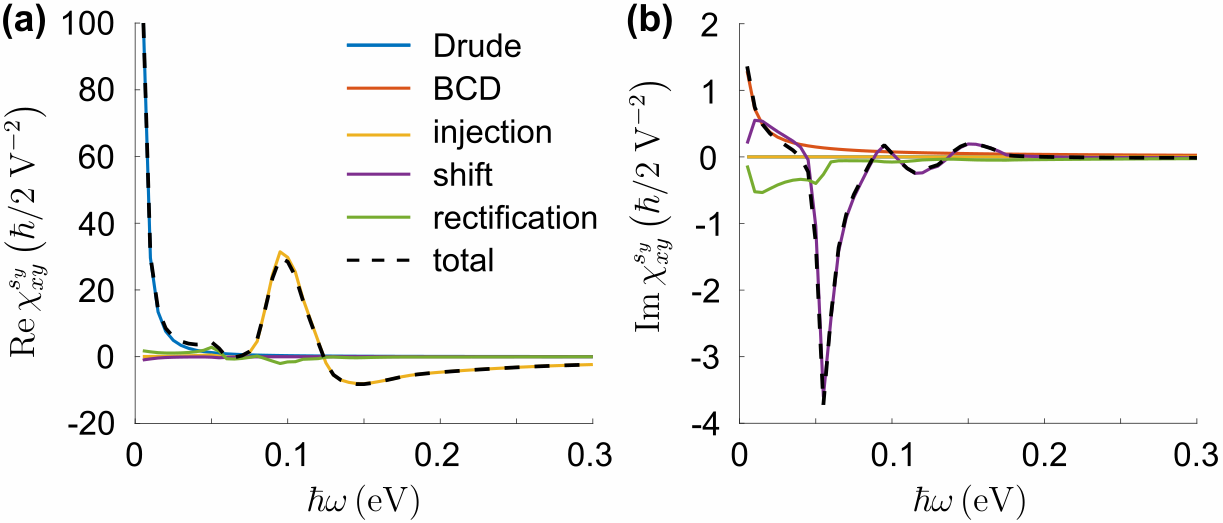}
  \caption{Numerical calculation of the nonzero nonlinear spin polarisation tensor $\Re \chi^{s_y}_{xy}$ (a) and $\Im \chi^{s_y}_{xy}$ (b) as a function of light frequency for prototypical hybrid parity magnet model. $\Re \chi^{s_y}_{xy}$ is dominated by the Drude contribution at low frequencies and the injection contribution at higher frequencies, while $\Im \chi^{s_y}_{xy}$ is dominated by the BCD contribution at low frequencies and shift contribution at higher frequencies. We have taken $\tau_s = 0.5$ ps as an illustrative spin relaxation time in calculating the injection contribution, see End Matter for details.
}
  \label{Fig:H3chiyxy}
\end{figure}

\textcolor{blue}{\it Coherent spin transport and domain parity in prototypical hybrid-parity magnet.}
As an illustration, we numerically calculate the nonlinear spin response tensors in a prototypical fully compensated 2D hybrid-parity magnet from Ref.~\cite{Luo2026} for a 2D honeycomb lattice. The details of the Hamiltonian are given in the End Matter. Our model preserves $g_A=\{\mathcal T U_z(\pi)\Vert I|0\}$, $g_P=\{U_y(\pi)\Vert P|0\}$, and $g_M=\{U_z(\pi)\Vert M_x|0\}$. The Bloch-spin texture has even $s_y(\bm k)$ and odd $s_z(\bm k)$ in $k$-space, with $s_x(\bm k)=0$. $g_M$ further requires $s_y(k_x,k_y) = -s_y (-k_x,k_y)$, leading to zero spin magnetization in equilibrium. The only nonzero nonlinear spin polarisation tensor is the magnetic-odd $\chi^{s_y}_{xy}$, and the allowed spin photocurrent tensors are $\sigma^{s_x}_{x,xx}$, $\sigma^{s_x}_{x,yy}$, and $\sigma^{s_x}_{y,xy}$ in the magnetic-odd sector, and $\sigma^{s_z}_{y,xx}$, $\sigma^{s_z}_{y,yy}$, and $\sigma^{s_z}_{x,xy}$ in the magnetic-even sector. Furthermore, the charge photocurrent is forbidden due to the presence of $g_P$, allowing for a pure spin photoresponse.

Fig.~\ref{Fig:H3chiyxy} shows the real and imaginary parts of $\chi^{s_y}_{xy}$ as a function of the light frequency. As discussed, $\chi^{s_y}_{xy}$ switches sign when the magnetic order is reversed, providing a diagnostic tool to probe magnetic domains. Furthermore, signatures of different microscopic contributions can be isolated by light frequency and polarisation dependence, since their real and imaginary parts are demanded to be magnetic even or odd by symmetry summarised in Table~\ref{tab:eta_scale}. The low frequency part of $\Re \chi^{s_y}_{xy}$ is dominated by the semiclassical Drude contribution, while interband transitions at around $\hbar \omega = 0.1$ eV gives rise to a resonant injection contribution. The imaginary (light helicity dependent) part, however, is dominated by the BCD contribution at low frequencies, and a resonant peak from shift contribution at around  $\hbar \omega = 0.05$ eV. In the clean limit $\eta \to 0$, $\Im\chi^{s_y}_{xy}=\Im\chi^{s_y,\mathrm{shift}}_{xy}+\OO{\eta}$, and the helicity dependent spin generation is determined by the resonant shift contribution. This allows for probing of different quantum geometric properties by optical means.

\begin{figure}[t]
  \includegraphics[width=\linewidth]{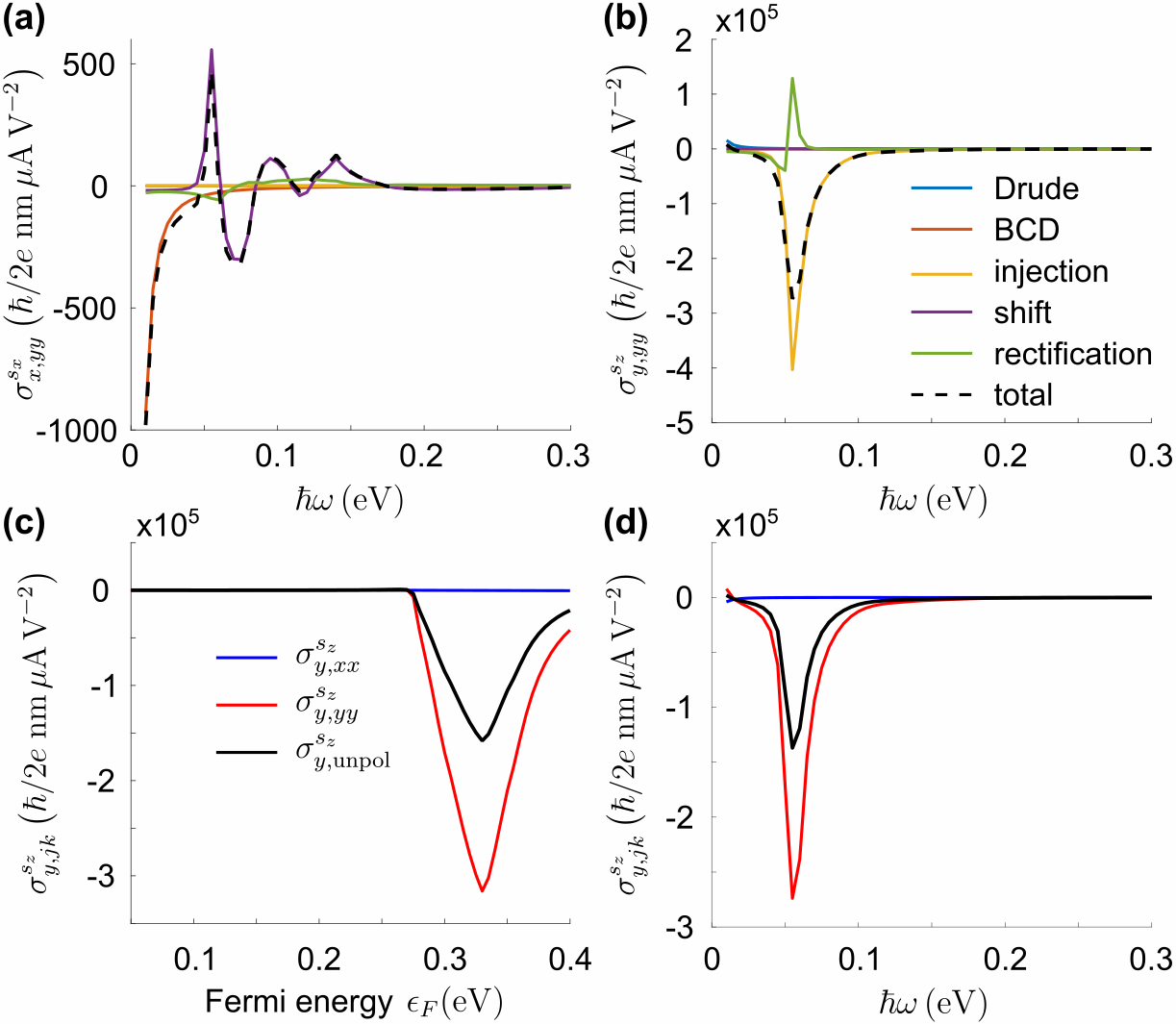}
  \caption{Numerical calculations of selected nonlinear spin photocurrent tensors for prototypical hybrid parity magnet model. (a) $\sigma^{s_x}_{x,yy}$ exhibits low frequency divergence due to the BCD contribution and resonant peaks due to the shift contribution. (b) $\sigma^{s_z}_{y,yy}$ shows prominent resonant peak due to the injection contribution. (c, d) Large spin photocurrent induced by unpolarised light can be tuned by Fermi energy (c) and irradiation frequency (d). We have taken $\tau_{j_s} = 0.2$ ps as an illustrative spin photocurrent relaxation time in calculating the injection contribution, see End Matter for details.
}
  \label{Fig:H3sigma}
\end{figure}

Similarly, the symmetry constraints allow us to distinguish the microscopic origins of the $\sigma^{s_a}_{i,jk}$ tensors by polarisation and frequency control, as illustrated in Fig.~\ref{Fig:H3sigma}. Fig.~\ref{Fig:H3sigma}(a) shows the real $\sigma^{s_x}_{x,yy}$ channel, where symmetry removes the magnetic-even Drude and injection contributions, leaving coherent BCD, shift, and rectification. This allows for detecting the magnetic domains by spin photocurrent measurement. In contrast, the real $\sigma^{s_z}_{y,yy}$ channel [Fig.~\ref{Fig:H3sigma}(b)] is magnetic even and exhibits a large resonant peak dominated by the injection contribution. 

\textcolor{blue}{\it Unpolarised spin transport.}
Materials lacking $P$ symmetry may host nonzero spin photocurrent $j^{s_a}_i$ induced by unpolarised light with $s_a$ parallel to any pure spin rotation symmetry axis and $i$ parallel to pure spatial rotation symmetry. For normally incident unpolarised light along $z$, $\langle E_jE_k^*\rangle=\mathcal I_E\delta_{jk}/2$ ($j,k=x,y$), where $\mathcal I_E=\langle|E_x|^2+|E_y|^2\rangle$. Hence the spin photocurrent averaged over polarisation is given by 
\begin{equation}
 \langle J_i^{s_a}\rangle=\mathcal I_E\sigma^{s_a}_{i,\mathrm{unpol}},\quad
 \sigma^{s_a}_{i,\mathrm{unpol}}=\frac{\sigma^{s_a}_{i,xx}+\sigma^{s_a}_{i,yy}}2.
 \label{eq:unpol}
\end{equation}
The hybrid-parity magnet model we consider here supports a large nonzero spin photocurrents $j^{s_x}_x$ and $j^{s_z}_y$ under unpolarised illumination while its charge photocurrent is forbidden. 
This is because the spin rotation in $g_P$ reverses $s_x$ and $s_z$, compensating the inversion sign of their current vertices. The mirror permits the optical trace for $j_x^{s_x}$ and $j_y^{s_z}$. Furthermore, $g_A$ demands $\sigma^{s_x}_{x{\rm unpol}}$ to be magnetic-odd, whereas $\sigma^{s_z}_{y, {\rm unpol}}$ to be magnetic-even, since $j^{s_x}_x$ reverses sign while $j^{s_z}_y$ remains unchanged under $g_A$ operation. 

Fig.~\ref{Fig:H3sigma}c and d show the numerical calculation of $\sigma^{s_z}_{y,{\rm unpol}}$, and its magnitude can be controlled by tuning Fermi energy and light frequency. An occupation-sensitive resonance at $\hbar \omega \approx 0.05$ eV of $\sigma^{s_z}_{y,\mathrm{unpol}}$ is observed with magnitude of order $10^5(\hbar/2e)\,\mu\mathrm A\,\mathrm{nm}\,\mathrm V^{-2}$ due to large injection contribution from optical transitions between Band 8 and 9. We emphasize that $\sigma^{s_z}_{y,{\rm unpol}}$ is magnetic even. This provides a pathway to polariser-free optical spin sources after averaging over time-reversed magnetic domains, whose response is governed by selectable microscopic processes.

In summary, our work provides a comprehensive mechanism-resolved symmetry analysis for second order spin polarisation and spin photocurrent in unconventional magnets. We demonstrate a practical scheme to probe spin group symmetry, magnetic domain and quantum geometry of unconventional magnets using these spin photoresponses and illustrate it in newly emerging hybrid- and odd-parity magnets. We also show large domain-even spin photocurrent induced by unpolarised light in prototypical hybrid-parity magnetic model. Our findings provide a pathway for all-optical material characterisation tools for unconventional magnets as well as polariser-free magnetic domain-insensitive spin source. 

{\it Acknowledgements.} This research is funded by the Singapore A*STAR IRG grant (M23M6c0102). 

\bibliography{refs.bib}

@article{Krempasky2024,
author={Krempask{\'y}, J.
and {\v{S}}mejkal, L.
and D'Souza, S. W.
and Hajlaoui, M.
and Springholz, G.
and Uhl{\'i}{\v{r}}ov{\'a}, K.
and Alarab, F.
and Constantinou, P. C.
and Strocov, V.
and Usanov, D.
and Pudelko, W. R.
and Gonz{\'a}lez-Hern{\'a}ndez, R.
and Birk Hellenes, A.
and Jansa, Z.
and Reichlov{\'a}, H.
and {\v{S}}ob{\'a}{\v{n}}, Z.
and Gonzalez Betancourt, R. D.
and Wadley, P.
and Sinova, J.
and Kriegner, D.
and Min{\'a}r, J.
and Dil, J. H.
and Jungwirth, T.},
title={Altermagnetic lifting of Kramers spin degeneracy},
journal={Nature},
year={2024},
month={Feb},
day={01},
volume={626},
number={7999},
pages={517-522},
issn={1476-4687},
doi={10.1038/s41586-023-06907-7},
url={https://doi.org/10.1038/s41586-023-06907-7}
}

@article{Wang2026,
author = {Wang, Jianhua and Yang, Xingyue and Yang, Zongmeng and Lu, Jing and Ho, Pin and Wang, Wenhong and Ang, Yee Sin and Cheng, Zhenxiang and Fang, Shibo},
title = {Pentagonal 2D Altermagnets: Material Screening and Altermagnetic Tunneling Junction Device Application},
journal = {Advanced Functional Materials},
volume = {36},
number = {28},
pages = {2505145},
doi = {https://doi.org/10.1002/adfm.202505145},
url = {https://advanced.onlinelibrary.wiley.com/doi/abs/10.1002/adfm.202505145},
year = {2026}
}

@misc{Feng2026,
      title={Crossed Electric Spin Generation in Multilayers}, 
      author={Xukun Feng and Wei Du and Jin Cao and Qian Niu and Shengyuan A. Yang and Cong Xiao},
      year={2026},
      eprint={2610.14668},
      archivePrefix={arXiv},
      primaryClass={cond-mat.mtrl-sci},
      url={https://arxiv.org/abs/2610.14668}, 
}

@article{Feng2025,
title = {Quantum metric nonlinear spin-orbit torque enhanced by topological bands},
journal = {Materials Today Quantum},
volume = {6},
pages = {100040},
year = {2025},
issn = {2950-2578},
doi = {https://doi.org/10.1016/j.mtquan.2025.100040},
url = {https://www.sciencedirect.com/science/article/pii/S2950257825000186},
author = {Xukun Feng and Weikang Wu and Hui Wang and Weibo Gao and Lay Kee Ang and Y.X. Zhao and Cong Xiao and Shengyuan A. Yang}
}

@article{Hayami2019,
  author = {Hayami, Satoru and Yanagi, Yuki and Kusunose, Hiroaki},
  title = {Momentum-Dependent Spin Splitting by Collinear Antiferromagnetic Ordering},
  journal = {J. Phys. Soc. Jpn.},
  volume = {88},
  pages = {123702},
  year = {2019},
  doi = {10.7566/JPSJ.88.123702},
  url = {https://doi.org/10.7566/JPSJ.88.123702}
}

@article{Smejkal2022Landscape,
  author = {{\v{S}}mejkal, Libor and Sinova, Jairo and Jungwirth, Tomas},
  title = {Emerging Research Landscape of Altermagnetism},
  journal = {Phys. Rev. X},
  volume = {12},
  pages = {040501},
  year = {2022},
  doi = {10.1103/PhysRevX.12.040501},
  url = {https://doi.org/10.1103/PhysRevX.12.040501}
}

@article{Jungwirth2026Signatures,
  author = {Jungwirth, Tomas and Sinova, Jairo and Fernandes, Rafael M.
    and Liu, Qihang and Watanabe, Hikaru and Murakami, Shuichi
    and Nakatsuji, Satoru and {\v{S}}mejkal, Libor},
  title = {Symmetry, microscopy and spectroscopy signatures of altermagnetism},
  journal = {Nature},
  volume = {649},
  pages = {837--847},
  year = {2026},
  doi = {10.1038/s41586-025-09883-2},
  url = {https://doi.org/10.1038/s41586-025-09883-2}
}

@Article{Tarasenko2005,
author={Tarasenko, S. A.
and Ivchenko, E. L.},
title={Pure spin photocurrents in low-dimensional structures},
journal={JETP Lett.},
year={2005},
month={Mar},
day={01},
volume={81},
number={5},
pages={231-235},
issn={1090-6487},
doi={10.1134/1.1921322},
url={https://doi.org/10.1134/1.1921322}
}

@article{deJuan2017,
author={de Juan, Fernando
and Grushin, Adolfo G.
and Morimoto, Takahiro
and Moore, Joel E.},
title={Quantized circular photogalvanic effect in Weyl semimetals},
journal={Nat. Commun.},
year={2017},
month={Jul},
day={06},
volume={8},
number={1},
pages={15995},
issn={2041-1723},
doi={10.1038/ncomms15995},
url={https://doi.org/10.1038/ncomms15995}
}

@article{Holder2020,
  title = {Consequences of time-reversal-symmetry breaking in the light-matter interaction: Berry curvature, quantum metric, and diabatic motion},
  author = {Holder, Tobias and Kaplan, Daniel and Yan, Binghai},
  journal = {Phys. Rev. Res.},
  volume = {2},
  issue = {3},
  pages = {033100},
  numpages = {18},
  year = {2020},
  month = {Jul},
  publisher = {American Physical Society},
  doi = {10.1103/PhysRevResearch.2.033100},
  url = {https://link.aps.org/doi/10.1103/PhysRevResearch.2.033100}
}

@MISC{Luo2026,
       author = {{Luo}, Xun-Jiang and {Li}, Dan and {Xiao}, Rui-Chun and {Shao}, Ding-Fu and {Li}, Lei and {Tian}, Mingliang and {Yao}, Yugui},
        title = "{Unconventional Magnetism: Symmetry Classification, Hybrid-parity and Unconstrained-parity Classes}",
         year = 2026,
        month = may,
          doi = {10.48550/arXiv.2605.21336},
archivePrefix = {arXiv},
       eprint = {2605.21336},
 primaryClass = {cond-mat.mtrl-sci},
       adsurl = {https://ui.adsabs.harvard.edu/abs/2026arXiv260521336L}
}

@misc{YeeSin2026,
      title={Hybrid-parity sliding multiferroics}, 
      author={Zhenzhou Guo and Jiangtao Yu and Shibo Fang and Jin Cao and Xiaodong Zhou and Yee Sin Ang and Wenhong Wang and Zhenxiang Cheng and Xiaotian Wang},
      year={2026},
      eprint={2607.24337},
      archivePrefix={arXiv},
      primaryClass={cond-mat.mtrl-sci},
      url={https://arxiv.org/abs/2607.24337}, 
}

@article{Smejkal2022,
  author = {{\v{S}}mejkal, Libor and Sinova, Jairo and Jungwirth, Tom{\'a}{\v{s}}},
  title = {Beyond Conventional Ferromagnetism and Antiferromagnetism: A Phase with Nonrelativistic Spin and Crystal Rotation Symmetry},
  journal = {Phys. Rev. X},
  volume = {12},
  pages = {031042},
  year = {2022},
  doi = {10.1103/PhysRevX.12.031042},
  url = {https://doi.org/10.1103/PhysRevX.12.031042}
}

@article{Yu2025,
  author = {Yu, Yue and Lyngby, Magnus B. and Shishidou, Tatsuya and Roig, Merc{\`e} and Kreisel, Andreas and Weinert, Michael and Andersen, Brian M. and Agterberg, Daniel F.},
  title = {Odd-Parity Magnetism Driven by Antiferromagnetic Exchange},
  journal = {Phys. Rev. Lett.},
  volume = {135},
  pages = {046701},
  year = {2025},
  doi = {10.1103/zk69-k6b2},
  eprint = {2501.02057},
  archivePrefix = {arXiv},
  primaryClass = {cond-mat.str-el},
  url = {https://doi.org/10.1103/zk69-k6b2}
}

@article{Song2025,
author={Song, Qian
and Stavri{\'{c}}, Srdjan
and Barone, Paolo
and Droghetti, Andrea
and Antonenko, Daniil S.
and Venderbos, J{\"o}rn W. F.
and Occhialini, Connor A.
and Ilyas, Batyr
and Erge{\c{c}}en, Emre
and Gedik, Nuh
and Cheong, Sang-Wook
and Fernandes, Rafael M.
and Picozzi, Silvia
and Comin, Riccardo},
title={Electrical switching of a p-wave magnet },
journal={Nature},
year={2025},
month={Jun},
day={01},
volume={642},
number={8066},
pages={64-70},
issn={1476-4687},
doi={10.1038/s41586-025-09034-7},
url={https://doi.org/10.1038/s41586-025-09034-7}
}

@article{Naka2019,
  author = {Naka, Makoto and Hayami, Satoru and Kusunose, Hiroaki
    and Yanagi, Yuki and Motome, Yukitoshi and Seo, Hitoshi},
  title = {Spin current generation in organic antiferromagnets},
  journal = {Nat. Commun.},
  volume = {10},
  pages = {4305},
  year = {2019},
  doi = {10.1038/s41467-019-12229-y},
  url = {https://doi.org/10.1038/s41467-019-12229-y}
}

@article{GonzalezHernandez2021,
  author = {Gonz{\'a}lez-Hern{\'a}ndez, Rafael and {\v{S}}mejkal, Libor
    and V{\'y}born{\'y}, Karel and Yahagi, Yuta and Sinova, Jairo
    and Jungwirth, Tom{\'a}{\v{s}} and {\v{Z}}elezn{\'y}, Jakub},
  title = {Efficient Electrical Spin Splitter Based on Nonrelativistic Collinear Antiferromagnetism},
  journal = {Phys. Rev. Lett.},
  volume = {126},
  pages = {127701},
  year = {2021},
  doi = {10.1103/PhysRevLett.126.127701},
  url = {https://doi.org/10.1103/PhysRevLett.126.127701}
}

@article{Smejkal2022GMR,
  author = {{\v{S}}mejkal, Libor and Birk Hellenes, Anna
    and Gonz{\'a}lez-Hern{\'a}ndez, Rafael and Sinova, Jairo
    and Jungwirth, Tomas},
  title = {Giant and Tunneling Magnetoresistance in Unconventional Collinear Antiferromagnets with Nonrelativistic Spin-Momentum Coupling},
  journal = {Phys. Rev. X},
  volume = {12},
  pages = {011028},
  year = {2022},
  doi = {10.1103/PhysRevX.12.011028},
  url = {https://doi.org/10.1103/PhysRevX.12.011028}
}

@article{Mencos2026,
author={Mencos, Jone
and Badura, Antonin
and Dolan, Eoin
and Beckert, Sebastian
and Gonz{\'a}lez-Hern{\'a}ndez, Rafael
and Sigales, Nicol{\'a}s
and Kokkeler, Tim
and Kounta, Isma{\"i}la
and Petit, Matthieu
and Guillemard, Charles
and Hellenes, Anna Birk
and Campos, Warlley
and Rial, Javier
and Kriegner, Dominik
and Baltz, Vincent
and Hueso, Luis E.
and Sinova, Jairo
and Bergeret, F. Sebastian
and Gomonay, Olena
and Jungwirth, Tom{\'a}{\v{s}}
and {\v{S}}mejkal, Libor
and Michez, Lisa
and Reichlova, Helena
and Casanova, F{\`e}lix},
title={Time-reversal-symmetry-breaking spin injection from a compensated magnet},
journal={Nature},
year={2026},
month={Nov},
day={01},
volume={659},
number={8139},
pages={69-75},
issn={1476-4687},
doi={10.1038/s41586-026-11227-7},
url={https://doi.org/10.1038/s41586-026-11227-7}
}

@article{Cheong2025,
author={Cheong, Sang-Wook
and Huang, Fei-Ting},
title={Altermagnetism classification},
journal={npj Quantum Materials},
year={2025},
month={Apr},
day={12},
volume={10},
number={1},
pages={38},
issn={2397-4648},
doi={10.1038/s41535-025-00756-5},
url={https://doi.org/10.1038/s41535-025-00756-5}
}

@article{Song2025Functional,
  author = {Song, Cheng and Bai, Hua and Zhou, Zhiyuan and Han, Lei
    and Reichlova, Helena and Dil, J. Hugo and Liu, Junwei
    and Chen, Xianzhe and Pan, Feng},
  title = {Altermagnets as a new class of functional materials},
  journal = {Nat. Rev. Mater.},
  volume = {10},
  pages = {473--485},
  year = {2025},
  doi = {10.1038/s41578-025-00779-1},
  url = {https://doi.org/10.1038/s41578-025-00779-1}
}

@article{Jungwirth2026Spintronics,
  author = {Jungwirth, T. and Sinova, J. and Wadley, P. and Kriegner, D.
    and Reichlov{\'a}, H. and Krizek, F. and Ohno, H.
    and {\v{S}}mejkal, L.},
  title = {Altermagnetic spintronics},
  journal = {Nat. Phys.},
  volume = {22},
  pages = {1012--1021},
  year = {2026},
  doi = {10.1038/s41567-026-03337-w},
  url = {https://doi.org/10.1038/s41567-026-03337-w}
}

@article{Xiao2022,
  title = {Intrinsic Nonlinear Electric Spin Generation in Centrosymmetric Magnets},
  author = {Xiao, Cong and Liu, Huiying and Wu, Weikang and Wang, Hui and Niu, Qian and Yang, Shengyuan A.},
  journal = {Phys. Rev. Lett.},
  volume = {129},
  issue = {8},
  pages = {086602},
  numpages = {6},
  year = {2022},
  month = {Aug},
  publisher = {American Physical Society},
  doi = {10.1103/PhysRevLett.129.086602},
  url = {https://link.aps.org/doi/10.1103/PhysRevLett.129.086602}
}

@article{Xu2021,
  title = {Light-induced static magnetization: Nonlinear Edelstein effect},
  author = {Xu, Haowei and Zhou, Jian and Wang, Hua and Li, Ju},
  journal = {Phys. Rev. B},
  volume = {103},
  issue = {20},
  pages = {205417},
  numpages = {11},
  year = {2021},
  month = {May},
  publisher = {American Physical Society},
  doi = {10.1103/PhysRevB.103.205417},
  url = {https://link.aps.org/doi/10.1103/PhysRevB.103.205417}
}

@article{Watanabe2021,
  title = {Chiral Photocurrent in Parity-Violating Magnet and Enhanced Response in Topological Antiferromagnet},
  author = {Watanabe, Hikaru and Yanase, Youichi},
  journal = {Phys. Rev. X},
  volume = {11},
  issue = {1},
  pages = {011001},
  numpages = {30},
  year = {2021},
  month = {Jan},
  publisher = {American Physical Society},
  doi = {10.1103/PhysRevX.11.011001},
  url = {https://link.aps.org/doi/10.1103/PhysRevX.11.011001}
}

@article{Ahn2020,
  title = {Low-Frequency Divergence and Quantum Geometry of the Bulk Photovoltaic Effect in Topological Semimetals},
  author = {Ahn, Junyeong and Guo, Guang-Yu and Nagaosa, Naoto},
  journal = {Phys. Rev. X},
  volume = {10},
  issue = {4},
  pages = {041041},
  numpages = {28},
  year = {2020},
  month = {Nov},
  publisher = {American Physical Society},
  doi = {10.1103/PhysRevX.10.041041},
  url = {https://link.aps.org/doi/10.1103/PhysRevX.10.041041}
}

@article{Xiong2022,
  title = {Polariton drag enabled quantum geometric photocurrents in high-symmetry materials},
  author = {Xiong, Ying and Shi, Li-kun and Song, Justin C. W.},
  journal = {Phys. Rev. B},
  volume = {106},
  issue = {20},
  pages = {205423},
  numpages = {6},
  year = {2022},
  month = {Nov},
  publisher = {American Physical Society},
  doi = {10.1103/PhysRevB.106.205423},
  url = {https://link.aps.org/doi/10.1103/PhysRevB.106.205423}
}

@article{Gao2026review,
  title = {Quantum geometry phenomena in condensed matter systems},
  author = {Gao, Anyuan and Nagaosa, Naoto and Ni, Ni and Xu, Su-Yang},
  journal = {Rev. Mod. Phys.},
  volume = {98},
  issue = {3},
  pages = {035005},
  numpages = {67},
  year = {2026},
  month = {Sep},
  publisher = {American Physical Society},
  doi = {10.1103/f78t-ky69},
  url = {https://link.aps.org/doi/10.1103/f78t-ky69}
}

@article{Dong2025,
  title = {Crystal symmetry selected pure spin photocurrent in altermagnetic insulators},
  author = {Dong, Ruizhi and Cao, Ranquan and Tan, Dian and Fei, Ruixiang},
  journal = {Phys. Rev. B},
  volume = {111},
  issue = {19},
  pages = {195210},
  numpages = {7},
  year = {2025},
  month = {May},
  publisher = {American Physical Society},
  doi = {10.1103/PhysRevB.111.195210},
  url = {https://link.aps.org/doi/10.1103/PhysRevB.111.195210}
}

@article{Sivianes2025,
  title = {Optical Signatures of Spin Symmetries in Unconventional Magnets},
  author = {Sivianes, Javier and Santos, Flaviano Jos{\'e} dos and Iba{\~n}ez-Azpiroz, Julen},
  journal = {Phys. Rev. Lett.},
  volume = {134},
  issue = {19},
  pages = {196907},
  numpages = {7},
  year = {2025},
  month = {May},
  publisher = {American Physical Society},
  doi = {10.1103/PhysRevLett.134.196907},
  url = {https://link.aps.org/doi/10.1103/PhysRevLett.134.196907}
}

@article{Merte2023,
    author = {Merte, M. and Freimuth, F. and Go, D. and Adamantopoulos, T. and Lux, F. R. and Plucinski, L. and Gomonay, O. and Blügel, S. and Mokrousov, Y.},
    title = {Photocurrents, inverse Faraday effect, and photospin Hall effect in Mn2Au},
    journal = {APL Materials},
    volume = {11},
    number = {7},
    pages = {071106},
    year = {2023},
    month = {07},
    issn = {2166-532X},
    doi = {10.1063/5.0149955},
    url = {https://doi.org/10.1063/5.0149955},
}

@article{Pei2026,
  title = {Switchable giant spin injection current in the Janus altermagnet ${\mathrm{Fe}}_{2}\mathrm{SSeO}$},
  author = {Pei, Fanxian and Li, Dan and Li, Lei and Lu, Jiajun and Zhang, Run-Wu and Yao, Yugui},
  journal = {Phys. Rev. B},
  volume = {114},
  issue = {10},
  pages = {105115},
  numpages = {9},
  year = {2026},
  month = {Aug},
  publisher = {American Physical Society},
  doi = {10.1103/9nlj-z5g7},
  url = {https://link.aps.org/doi/10.1103/9nlj-z5g7}
}

@misc{Rappe2026,
      title={Reading Altermagnetic Domains with Photon Drag}, 
      author={Sayed Ali Akbar Ghorashi and Andrew M. Rappe},
      year={2026},
      eprint={2610.10970},
      archivePrefix={arXiv},
      primaryClass={cond-mat.mes-hall},
      url={https://arxiv.org/abs/2610.10970}, 
}

@article{Chakraborty2025,
  author = {Chakraborty, Atasi and {Birk Hellenes}, Anna and Jaeschke-Ubiergo, Rodrigo and Jungwirth, Tom{\'a}{\v{s}} and {\v{S}}mejkal, Libor and Sinova, Jairo},
  title = {Highly efficient non-relativistic {Edelstein} effect in nodal {$p$}-wave magnets},
  journal = {Nature Communications},
  volume = {16},
  pages = {7270},
  year = {2025},
  doi = {10.1038/s41467-025-62516-0},
  url = {https://www.nature.com/articles/s41467-025-62516-0}
}

@article{Lihm2022,
  author = {Lihm, Jae-Mo and Park, Cheol-Hwan},
  title = {Comprehensive theory of second-order spin photocurrents},
  journal = {Phys. Rev. B},
  volume = {105},
  pages = {045201},
  year = {2022},
  doi = {10.1103/PhysRevB.105.045201},
  url = {https://doi.org/10.1103/PhysRevB.105.045201}
}

@article{Xiao2026,
  title = {DC spin current driven by low-frequency light in altermagnets},
  author = {Xiao, Yihua and Dong, Ruizhi and Wang, Xiaomu and Fei, Ruixiang},
  journal = {Phys. Rev. B},
  volume = {114},
  issue = {1},
  pages = {014405},
  numpages = {8},
  year = {2026},
  month = {Jul},
  publisher = {American Physical Society},
  doi = {10.1103/3j6x-397h},
  url = {https://link.aps.org/doi/10.1103/3j6x-397h}
}

@article{Sarkar2024,
  author = {Sarkar, Sayan and Das, Sunit and Mandal, Debottam and Agarwal, Amit},
  title = {Light-induced nonlinear resonant spin magnetization},
  journal = {New J. Phys.},
  volume = {27},
  number = {11},
  pages = {113503},
  year = {2025},
  eprint = {2409.12142},
  archivePrefix = {arXiv},
  primaryClass = {cond-mat.mes-hall},
  doi = {10.1088/1367-2630/ae1868},
  url = {https://doi.org/10.1088/1367-2630/ae1868}
}

@Article{Ma2023,
author={Ma, Qiong
and Krishna Kumar, Roshan
and Xu, Su-Yang
and Koppens, Frank H. L.
and Song, Justin C. W.},
title={Photocurrent as a multiphysics diagnostic of quantum materials},
journal={Nature Reviews Physics},
year={2023},
month={Mar},
day={01},
volume={5},
number={3},
pages={170-184},
issn={2522-5820},
doi={10.1038/s42254-022-00551-2},
url={https://doi.org/10.1038/s42254-022-00551-2}
}

@article{Takagi2026,
  title = {Quantum geometry in low-energy linear and nonlinear optical responses of the magnetic Rashba semiconductor (Ge,Mn)Te},
  author = {Takagi, Tsubasa and Watanabe, Hikaru and Yoshimi, Ryutaro and Sato, Yuki and Toyoda, Shingo and Tsukazaki, Atsushi and Takahashi, Kei S. and Kawasaki, Masashi and Tokura, Yoshinori and Ogawa, Naoki},
  journal = {Phys. Rev. B},
  volume = {113},
  issue = {10},
  pages = {104417},
  numpages = {7},
  year = {2026},
  month = {Mar},
  publisher = {American Physical Society},
  doi = {10.1103/dp31-y358},
  url = {https://link.aps.org/doi/10.1103/dp31-y358}
}

@article{Sodemann2015,
  author = {Sodemann, Inti and Fu, Liang},
  title = {Quantum Nonlinear Hall Effect Induced by Berry Curvature Dipole in Time-Reversal Invariant Materials},
  journal = {Phys. Rev. Lett.},
  volume = {115},
  pages = {216806},
  year = {2015},
  doi = {10.1103/PhysRevLett.115.216806},
  url = {https://doi.org/10.1103/PhysRevLett.115.216806}
}

@Article{Guo2026,
author={Guo, Tianxia
and Gao, Yinlu
and Zhao, Jijun
and Jiang, Xue},
title={Ultrafast spin dynamics in altermagnetic Cr2Se2O monolayer},
journal={npj Computational Materials},
year={2026},
month={May},
day={08},
volume={12},
number={1},
pages={241},
issn={2057-3960},
doi={10.1038/s41524-026-02108-8},
url={https://doi.org/10.1038/s41524-026-02108-8}
}

@article{Li2025spintau,
    author = {Li, Peiyan and Wu, Na and Liu, Shanshan and Cheng, Yu and Gong, Piming and Tong, Junwei and Liu, Jianan and He, Wei and Xiu, Faxian and Zhao, Jimin and Meng, Sheng and Wu, Xiaojun},
    title = {Above-Curie-temperature ultrafast terahertz emission and spin current generation in a 2D superlattice (Fe3GeTe2/CrSb)3},
    journal = {National Science Review},
    volume = {12},
    number = {3},
    pages = {nwae447},
    year = {2025},
    month = {03},
    issn = {2095-5138},
    doi = {10.1093/nsr/nwae447},
    url = {https://doi.org/10.1093/nsr/nwae447},
}

@Article{Dai2024,
author={Dai, Yudi
and Xiong, Junlin
and Ge, Yanfeng
and Cheng, Bin
and Wang, Lizheng
and Wang, Pengfei
and Liu, Zenglin
and Yan, Shengnan
and Zhang, Cuiwei
and Xu, Xianghan
and Shi, Youguo
and Cheong, Sang-Wook
and Xiao, Cong
and Yang, Shengyuan A.
and Liang, Shi-Jun
and Miao, Feng},
title={Interfacial magnetic spin Hall effect in van der Waals Fe3GeTe2/MoTe2 heterostructure},
journal={Nature Communications},
year={2024},
month={Feb},
day={07},
volume={15},
number={1},
pages={1129},
issn={2041-1723},
doi={10.1038/s41467-024-45318-8},
url={https://doi.org/10.1038/s41467-024-45318-8}
}

@misc{Bai2026,
      title={Time-reversal-odd Transport in Odd-Parity Magnets}, 
      author={Ling Bai and Si Li and Libor Šmejkal and Yugui Yao and Wanxiang Feng},
      year={2026},
      eprint={2609.25919},
      archivePrefix={arXiv},
      primaryClass={cond-mat.mtrl-sci},
      url={https://arxiv.org/abs/2609.25919}, 
}

@article{Chen2024,
  title = {Enumeration and Representation Theory of Spin Space Groups},
  author = {Chen, Xiaobing and Ren, Jun and Zhu, Yanzhou and Yu, Yutong and Zhang, Ao and Liu, Pengfei and Li, Jiayu and Liu, Yuntian and Li, Caiheng and Liu, Qihang},
  journal = {Phys. Rev. X},
  volume = {14},
  issue = {3},
  pages = {031038},
  numpages = {33},
  year = {2024},
  month = {Aug},
  publisher = {American Physical Society},
  doi = {10.1103/PhysRevX.14.031038},
  url = {https://link.aps.org/doi/10.1103/PhysRevX.14.031038}
}

\clearpage

\onecolumngrid

\smallskip

\begin{center}
\textbf{\large End Matter}
\end{center}

\twocolumngrid

\section{Appendix A: Microscopic formulas and relaxation convention}
Nonlinear responses can be derived by coupling a Bloch Hamiltonian $H_0 (\vec k)$ to an external AC electric field $\vec E (t) = \frac{1}{2} \vec E e^{-i\omega t} +c.c$ via dipole interaction $H'(t) = e\vec E (t) \cdot \hat{\vec r}$ in the length gauge. Here $\vec r_{mn} = i\nabla_{\vec k} \delta_{mn} + \vec R_{mn} $ is the matrix element of the position operator, $\vec R_{mn} = i \la u_m| \nabla_{\vec k} | u_n \ra$ is the Berry connection, and $|u_n\ra$ is the Bloch state. The nonequilibrium density matrix up to each order of electric field can be solved by the quantum Liouville equation $i\hbar \p_t \rho = [H, \rho]$, where $H = H_0(\vec k) + H'(t)$ is the perturbed Hamiltonian. The first order density matrix is 
\be
\rho^{(1)}_{mn} (\omega) = \sum_j {e E_j (\omega)} \frac{-i\p_j f_n \delta_{mn} + R^j_{mn} F_{mn}}{\epsilon_{mn} - \hbar \omega - i \eta}.
\ee
The time independent second order density matrix is
\begin{align}
&\rho^{(2), {\rm DC}}_{mn} = -\sum_{j,\omega_1= \pm \omega} \frac{e E_j (\omega_1)}{\Lambda_{mn}} \times \nonumber \\
&\quad\left[ i \p_j \rho^{(1)}_{mn} (-\omega_1) + \sum_l R^j_{ml} \rho^{(1)}_{ln} (-\omega_1) - \rho^{(1)}_{ml} (-\omega_1) R^j_{ln} \right] 
\end{align}
We have defined $\Omega=\hbar\omega$, $\epsilon_{mn}=\epsilon_m-\epsilon_n$, $F_{mn}=f_m-f_n$, $X_{mn}=\epsilon_{mn}+\Omega$, and $\Lambda_{mn}=\epsilon_{mn}-2i\eta$. Here $f_n$ is the Fermi distribution and $\partial_j=\partial/\partial k_j$. We also define $\vec E(\omega) = \vec E/2$ and $E(-\omega) = \vec E^*/2$. For an observable $\hat O$, the second order DC response is given by $O^{(2),{\rm DC}} = \Tr[\rho^{(2), {\rm DC}} O ]$. 

We focus on the second order DC spin polarisation $\hat O=\hat s_a$ and spin current $\hat O  = \hat j_i^{s_a}$. Table~\ref{tab:microscopic} gives the five contributions to $K^O_{jk} $ where $K^{s_a} = \chi$ and $K^{j^{s_a}}=\sigma$, respectively. We separate the responses into Fermi surface terms, resonant optical transitions and virtual interband terms. The Drude term is semiclassical and comes from the single band Fermi surface redistribution $\delta_{mn}$. The BCD term arises from the Fermi surface dependent interband coherences from $\p_j\rho^{(1)}_{mn}$. To obtain its expression in Table~\ref{tab:microscopic}, we have defined $d_\pm=\pm\Omega-i\eta$. The pole of the energy denominator $\lim_{\eta \to 0} \frac{1}{X_{mn} - i \eta} = \mathcal P (1/X_{mn})  + i \pi \delta(X_{mn})$ gives the principal part (virtual interband processes) and the resonant part (on-shell optical transitions). The injection and shift contributions arise from as the resonant terms coming from the on-diagonal and off-diagonal components of the second order density matrix $\rho^{(2), {\rm DC}}$. In particular, the injection contribution is typically written as $O^{\rm inj} = \tau^{\rm inj} \p_t \sum_n O_{nn} \rho^{(2),{\rm DC}}_{nn, {\rm resonant}}$, where $\tau^{\rm inj}$ is the accumulation/relaxation time of the change in expectation value of $\hat O$ due to optical transitions. See a detailed discussion in Appendix C. Lastly, the rectification term collects the principal parts of the interband contributions.

To simplify the notation, we introduce the off-diagonal response matrix
\begin{equation}
 Q^O_{nm}(\eta)=-\frac{i\hbar O_{nm}}{\epsilon_{mn}-2i\eta} (1-\delta_{mn}),
 \label{eq:Q}
\end{equation}
and its full generalized derivative
\begin{align}
 (\mathcal D_jQ^O)_{nm}
 &=\big[\partial_j-i(R^j_{nn}-R^j_{mm})\big]Q^O_{nm}\nonumber\\
 &\quad+i\sum_{\ell\ne n,m}(Q^O_{n\ell}R^j_{\ell m}-R^j_{n\ell}Q^O_{\ell m}).
 \label{eq:DQ}
\end{align}
For brevity, we define $\mathcal Q^O_{j;nm}(\eta)=(\mathcal D_jQ^O(\eta))_{nm}$, and
\begin{align}
 W^O_{jk,mn}&=F_{mn}(O_{nn}-O_{mm})R^k_{mn}R^j_{nm},\label{eq:W}\\
 S^O_{jk,mn}&=F_{mn}\big[R^k_{mn}\mathcal Q^O_{j;nm}(0)\nonumber\\
 &\hspace{24mm}-R^j_{nm}\mathcal Q^O_{k;mn}(0)\big],\label{eq:S}\\
 G^O_{jk,mn}(\eta)&=F_{mn}\big[R^k_{mn}\mathcal Q^O_{j;nm}(\eta)\nonumber\\
 &\hspace{24mm}+R^j_{nm}\mathcal Q^O_{k;mn}(\eta)\big],\label{eq:G}\\
 P_\eta(X)&=\frac{X}{X^2+\eta^2},\quad
 H_\eta(X)=\frac{X^2-\eta^2}{(X^2+\eta^2)^2}.
 \label{eq:principal}
\end{align}
For small $\eta$, we have $P_\eta\to\mathcal P(1/X)$ and $H_\eta\to\operatorname{Pf}(1/X^2)$ the principal part of the interband process, where Pf is the Hadamard finite part.

\section{Appendix B: Magnetic parity of spin photoresponse tensors}
In this section, we analyse magnetic parity dependence of the spin photoresponse tensors by considering how each matrix element transforms under time reversal $\mathcal T$. Upon time reversal, the bare Bloch Hamiltonian satisfies 
\be
\mathcal T H_0^{\mathcal M} (\vec k) \mathcal T^{-1} = H_0^{-\mathcal M} (-\vec k).
\ee
By considering $H_0^{-\mathcal M} (-\vec k) \mathcal T |u_n^{\mathcal M} (\vec k)\ra = \epsilon_n^{\mathcal M}(\vec k) \mathcal T |u_n^{\mathcal M} (\vec k)\ra$, we obtain 
\be\label{eq:E_state}
\epsilon_{\bar n}^{-\mathcal M} (-\vec k) = \epsilon_n^{\mathcal M} (\vec k), \quad |u_{\bar n}^{-\mathcal M} (-\vec k) = e^{i \phi_n(\vec k)} \mathcal T |u_n^{\mathcal M} (\vec k)\ra,
\ee
where band $\bar n$ is the time reversal partner band of band $n$, and $\phi_n (\vec k)$ is a phase associated with the transformation. This sewing phase cancels out, and thus we will suppress it subsequently. Using Eq.~\ref{eq:E_state} and noting that $\mathcal T \hat{\vec s} \mathcal T^{-1} = -\hat{\vec s}$ and $\mathcal T \hat{\vec j}^{s_a} \mathcal T^{-1} = \hat{\vec j}^{s_a}$, we obtain the relations of the relevant matrix elements with its time-reversal partners: 
\begin{align}\label{eq:matrix_T}
&\vec R_{mn}^{\mathcal M} (\vec k) = \left[\vec R_{mn}^{-\mathcal M} (-\vec k) \right]^*, \quad \vec s_{mn}^{\mathcal M} (\vec k) = - \left[\vec s_{mn}^{-\mathcal M} (-\vec k) \right]^* ,\nonumber \\
&\vec j_{mn}^{s_a,\mathcal M} (\vec k) = \left[\vec j_{mn}^{s_a, -\mathcal M} (-\vec k) \right]^*.
\end{align}
Substituting Eq.~\ref{eq:matrix_T} into the expressions in Table~\ref{tab:microscopic}, we obtain
\begin{align}
 K^{O,X}[-\mathcal M;\eta]&=\zeta_O\big[K^{O,X}[\mathcal M;-\eta]\big]^*,\nonumber\\[-2pt]
 &\hspace{13mm}X=\mathrm D,\mathrm{BCD},\mathrm{rec},\label{eq:parity_nonres}\\
 K^{O,\mathrm{inj}}[-\mathcal M]&=\zeta_O\big[K^{O,\mathrm{inj}}[\mathcal M]\big]^*,\\
 K^{O,\mathrm{shift}}[-\mathcal M]&=-\zeta_O\big[K^{O,\mathrm{shift}}[\mathcal M]\big]^*.
\end{align}
Here, $\zeta_O=-1$ and $+1$ for $\chi^{s_a}_{jk}$ and $\sigma^{s_a}_{i,jk}$, respectively. By applying $K^{(E/O)}[\mathcal M]=\{K[\mathcal M]\pm K[-\mathcal M]\}/2$, we find that the magnetic even and odd components exhibit different dissipation dependence, as shown in Table~\ref{tab:eta_scale}.

\section{Appendix C: Accumulation time for injection contribution}
The injection contribution arises from the difference in electron spin and spin current expectation values when it undergoes interband transition, and it can be written as $\chi^{s_a, \rm inj}_{jk} = \tau^{\rm inj}_s \tilde \chi^{s_a, \rm inj}_{jk} $ and $\sigma^{s_a, \rm inj}_{i,jk} = \tau^{\rm inj}_{j_s} \tilde \sigma^{s_a, \rm inj}_{i,jk} $, where $\tilde \chi^{s_a, \rm inj}_{jk}$ and $\tilde \sigma^{s_a, \rm inj}_{i,jk} $ are the injection rates, and $\tau_{s(j_s)}^{\rm inj}$ is the effective time over which the injection spin (spin photocurrent)
relaxes/accumulates. For continuous wave excitation, $\tau^{\rm inj}_s =\tau_s$ and $\tau^{\rm inj}_{j_s} =\tau_{j_s}$ are the optically excited spin imbalance and spin current relaxation times. We note that $\tau_{j_s}$ are not necessarily equal to momentum relaxation or charge transport times, since electron–electron collisions can relax the spin current even when they do not relax the ordinary charge current~\cite{Tarasenko2005}. For ultrafast excitations with pulse width much smaller than the relevant relaxation times, $\tau_{s(j_s)}^{\rm inj}$ can be taken as the pulse width~\cite{deJuan2017,Holder2020}.

\section{Appendix D: Prototypical 2D hybrid parity magnet}
In this section, we present a 2D hybrid parity magnet on a honeycomb lattice from Ref.~\cite{Luo2026}. The tight binding Hamiltonian is~\cite{Luo2026}:
\be
H = \sum_{\la ij\ra s} t c^\dagger_{i, s} c_{j,s} + J \sum_{i, s, s'} \vec m_i \cdot c^\dagger_{i, s} \hat{\vec s}_{s, s'} c_{i,s'},
\ee
where $t$ is the nearest-neighbour hopping parameter, $J$ is the exchange coupling strength, and $\vec m_i$ is a unit vector specifying the orientation of the local magnetic moment on site $i \in \{ a, b, c,d,e,f\}$ arranged in a counterclockwise direction on the vertices of the honeycomb lattice. As an illustration, we follow Ref~\cite{Luo2026} and take $t=1$ eV, $J = 0.5$ eV, $\vec m_a = (1/2, -\sqrt{3}/2,0)$, $\vec m_b = (-1/2, \sqrt{3}/2,0)$, $\vec m_d = -(1/2, \sqrt{3}/2,0)$, $\vec m_e = (1/2, \sqrt{3}/2,0)$, and $\vec m_c = \vec m_f = 0$. We further assume Fermi energy $\epsilon_F = 0.3$ eV comparable to Ref.~\cite{Luo2026}, a nearest-neighbour hopping distance of 0.3 nm and $\eta = 2$ meV, corresponding to a relaxation time of around 0.3 ps. For injection contribution of the nonlinear spin polarisation, we take $\tau^{\rm inj}_{s} = \tau_{s} = 0.5$ ps as an estimate for typical spin relaxation time in 2D magnet~\cite{Li2025spintau, Guo2026}, and $\tau^{\rm inj}_{j_s} = \tau_{j_s} = 0.2$ ps for an estimate of typical spin current relaxation time~\cite{Dai2024}.

\begin{figure}[t]
  \includegraphics[width=\linewidth]{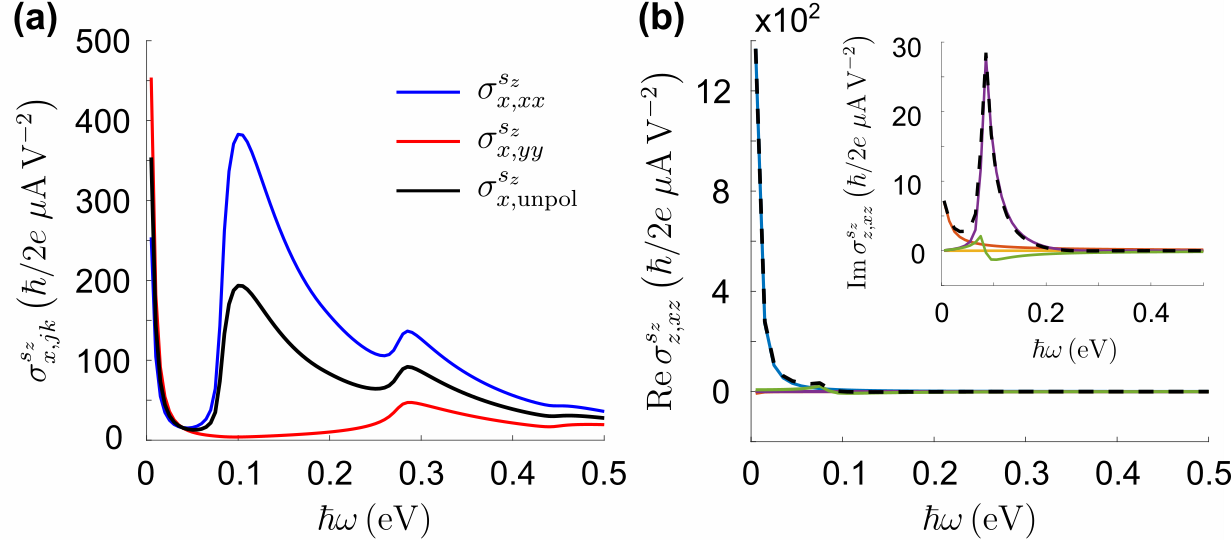}
  \caption{Numerical calculations of CeNiAsO-inspired odd parity magnet model. (a) $\sigma^{s_z}_{x,{\rm unpol}}$ vs frequency graph shows sizable spin photocurrent induced by normal (along $z$-direction) incidence of unpolarised light. (b) The real and imaginary parts of $\sigma^{s_z}_{z,xz}$ as a function of light frequency. We have assumed $\eta =5$ meV and $\tau^{\rm inj}_{j_s} = 0.2$ ps for a numerical illustration. 
}
  \label{Fig:CeNiAsO}
\end{figure}

\section{Appendix E: Nonlinear spin responses in odd parity magnet model.}

We present numerical calculations of spin photocurrent tensors in a CeNiAsO-inspired eight-band model of a coplanar odd parity magnet from Ref.~\cite{Yu2025}, which provides a contrasting example. The combination of $\{U_z(\pi)\Vert I|\bm d\}$ and $\{\mathcal T U_z(\pi)\Vert I|0\}$ permits only $s_z (\vec k) = -s_z( - \vec k)$ to be odd in the $k$-space. The additional $\{U_x(\pi)\Vert P|0\}$ forbids every homogeneous second-order spin-polarisation component, while allowing $s_z$ spin current. The surviving current is magnetic-even; a spin-independent $y$ mirror requires an even number of $y$ indices among $i,j,k$. Thus the model illustrates symmetry-enforced separation of spin transport from uniform spin polarisation, without conservation of physical $s_z$.


Unpolarised light can consequently induce $J_x^{s_z}$, as illustrated by the frequency-dependent spectrum in Fig.~\ref{Fig:CeNiAsO}a. 
Furthermore, an additional scalar out-of-plane dispersion adopted in the model, $H=H_\parallel+h_z(k_z)I$~\cite{Yu2025}, eliminates direct $z$-polarised interband vertices but still permits mixed-channel spin-current coherence. In particular, $\sigma^{s_z}_{z,xz}$ can contain shift and rectification contributions (Fig.~\ref{Fig:CeNiAsO}b inset) even though its population injection vanishes, as illustrated by Fig.~\ref{Fig:CeNiAsO}b. Thus, $\sigma^{s_z}_{z,xz}$ distinguishes optical excitation vertices from the measured interband spin-current vertex.



\end{document}